\documentclass[aps,prl,twocolumn,showpacs,nofootinbib]{revtex4}
\usepackage{graphicx} \usepackage{amsmath} \usepackage{amssymb}
\usepackage{amsfonts} \usepackage{bm}

\begin{document}

\newcommand{\be}{\begin{equation}} \newcommand{\ee}{\end{equation}}
\newcommand{\bea}{\begin{eqnarray}}\newcommand{\eea}{\end{eqnarray}}

\title{Scalar field dynamics in  warped $\rm{AdS}_3$ black hole background}

\author{Sayan K. Chakrabarti} \email{sayan.chakrabarti@saha.ac.in}

\author{Pulak Ranjan Giri} \email{pulakranjan.giri@saha.ac.in}

\author{Kumar S. Gupta} \email{kumars.gupta@saha.ac.in}

\affiliation{Theory Division, Saha Institute of Nuclear Physics,
1/AF Bidhannagar, Calcutta 700064, India}

\begin{abstract}
We study the normal modes of a scalar field in the background of a warped  $\rm{AdS}_3$ black
hole which arises in topologically massive gravity. We discuss the normal mode spectrum using the brick wall boundary condition. In addition, we investigate the possibility of a more general boundary condition for the scalar field.
\end{abstract}

\pacs{04.70.Dy, 04.70.-s, 04.60.Kz, 03.65.Ge}

\date{\today}

\maketitle

\section{INTRODUCTION}

Topologically massive gravity in $2+1$ dimensions with a negative cosmological constant $-1/l^2$ is of much interest nowadays. It admits an $\rm{AdS}_3$ vacuum solution for arbitrary value of the graviton mass $\tilde\mu$. It was conjectured in \cite{strom1} that a consistent quantum theory of chiral gravity can be defined at $\tilde\mu l = 1$. There are unstable and inconsistent vacua in this system when $\tilde\mu l \ne 1$. For $\tilde\mu l >1$ this happens due to the presence of massive graviton with negative energy in the bulk, whereas for $\tilde \mu l<1$ it is due to negative energy of the BTZ black hole \cite{strom1}. However it was found that for $\tilde\mu l\ne 1$, one can not exclude the possibility that these theories have other stable ground states. In fact a warped $\rm{AdS}_3$ vacua has been found for every $\tilde\mu$ \cite{vuorio, vuorio1}. The warped $\rm{AdS}_3$ geometry is viewed as a fibration of the real line with a constant warp factor over $\rm{AdS}_2$ 
 \cite{beng1,henn1,beng2}. The isometry $SL(2,\mathbb{R})_L\times SL(2,\mathbb{R})_R$ of $\rm{AdS}_3$ then breaks down to $SL(2,\mathbb{R})\times U(1)$. One of the solutions free from the naked closed timelike curves is the spacelike stretched black hole solution \cite{strom2,nutku,clement,clement1,clement2}. In \cite{strom2}, it was shown that all the above mentioned solutions are locally equivalent and space like warped $\rm{AdS}_3$ is a stable vacuum of a quantum theory of gravity at $\tilde\mu l >3$.

In order to study the properties of the warped $\rm{AdS}_3$ black hole it is useful to analyze the dynamics of a scalar field in this background. Such an approach has been used as a simple probe to study the properties of a black hole, including the near-horizon conformal structure \cite{ksg1} and black hole entropy \cite{thooft1,sorkin,suss,satoh,thooft2,thooft3,ksg2,ksg3,ksg4}. Such an approach to study the black hole geometry is not without problems. The scalar field is typically not well behaved at the horizon and the free energy may show divergence due to infinite number of modes contributing to it near the horizon \cite{thooft1,sorkin,suss,satoh,thooft2,thooft3}. Moreover, the $\rm{AdS}_3$ background is not globally hyperbolic, thereby leading to difficulties in predicting the field propagation \cite{isham,wald}. Many of these issues can be addressed
by imposing suitable boundary conditions on the scalar fields. For example, in the brick wall approach, one assumes that the scalar field vanishes at a certain distance away from the horizon, thereby leading to a well defined prescription for finding out the corresponding free energy and entropy \cite{thooft1,sorkin}. 

On the other hand a vanishing boundary condition on the scalar field at spatial infinity provides a way to define unitary time evolution in $\rm{AdS}_3$.

It is however possible to put an infrared cutoff at a large distance $R$ as was done in the original discussion of the black hole entropy using a brick wall cutoff in the paper by 't Hooft  \cite{thooft1}. However, as it was shown in that paper, only the modes in the near-horizon region are responsible for the Bekenstein-Hawking entropy of the back hole, that is why we  focus  on the boundary condition near the horizon and  take the infrared cutoff $R$ effectively to infinity. Also, since $\rm{AdS}_3$ is not globally hyperbolic, it is important to make the scalar field vanish at infinity, which leads to a well defined time evolution. This is another reason for our choice of the condition at infinity.
For a discussion of the boundary condition at spatial infinity in $\rm{AdS}_3$ see  Ref. \cite{suss,ksg4}.

The physical information that can be obtained by the scalar field analysis is often strongly dependent on the choice of boundary conditions \cite{satoh,ken1,ken2}.

For example, when the entropy of a black hole is calculated using a scalar field, the results depend critically on the boundary conditions used. Typically the brick wall boundary condition is used in the literature. What we wish to point out here is that the brick wall boundary condition used in the literature is by no means unique and there exists a whole new class of other boundary conditions which are equally well admissible by the criterion of self-adjointness or unitary evolution. While addressing quantum aspects of gravity, we have no a priori idea of what the correct boundary condition should be. 

It is therefore useful to classify all possible allowed boundary conditions which would lead to a well defined time evolution of the scalar field and to investigate the dependence of the physical quantities on such boundary conditions. A preliminary step in this direction for the BTZ black hole \cite{btz} was taken by us in \cite{cgg}.

In this paper, we shall study the normal modes of the scalar field in the background of a warped $\rm{AdS}_3$ black hole. The analysis of scalar fields in this geometry has already been used to study the quasinormal modes \cite{qnm}, absorption cross sections \cite{abscross} and brick wall entropy \cite{brckwall}. We start with the assumption of a vanishing brick wall type boundary condition for the scalar field near the outer horizon and find the normal modes. We shall show that modes with real eigenvalues that are square integrable at infinity exist only for certain range of the energy eigenvalues, or equivalently, for certain ranges of the quantum numbers. This behavior is quite different from that for the BTZ, for which the normal modes with brick wall type boundary condition exist without any such bound on the energy.

Next we shall try to obtain the most general boundary conditions which the scalar field can obey and yet have a well defined time evolution. In this process, we shall be guided by the principle of self-adjointness \cite{reed} and will find all possible self-adjoint extensions of the radial part of the scalar field Hamiltonian. For this, we shall first use the method of deficiency indices due to von Neumann \cite{reed} to classify the various boundary
conditions that can be imposed on the scalar field. We show that for
certain range of the system parameters, there exists a one parameter
family of self-adjoint extensions of the corresponding Klein-Gordon
equation. There are however certain subtleties associated with this
procedure. For this reason, we shall try to find the inequivalent
spectra of the problem using a more physical approach \cite{wilc,cgg}.


This paper is organized as follows. In Section 2 we set up the
scalar field propagation problem in the spacelike stretched black hole space-time in warped $\rm{AdS}_3$ background and analyze the normal mode using brick wall boundary condition with a very brief account of the calculation of scalar field entropy following the methods used by Ichinose and Satoh in \cite{satoh}. In Section 3
we use von Neumann's method of self-adjoint extension to find the
deficiency indices of the radial part of the Klein-Gordon operator.
 Section 4 concludes the paper with some
discussions and an outlook.

\section{Normal modes with brick wall boundary condition}
We start with a brief review of the black hole solutions in topologically massive gravity with negative cosmological constant. The action for topologically massive gravity
\cite{deser1, deser2,carlip} with a negative cosmological constant is given by
\begin{eqnarray}
\nonumber I_{TMG}&=&\frac{1}{16\pi G}\int d^3x\sqrt{-g}(R+\frac{2}{l^2})+\\
\nonumber &&\frac{l}{96\pi G\nu}\int d^3x \sqrt{-g}\epsilon^{\lambda\mu\nu}\Gamma^{\alpha}_{\lambda\sigma}\times\\
&&\left(\partial_{\mu}\Gamma^{\sigma}_{\alpha\nu}+\frac{2}{3}\Gamma^{\sigma}_{\mu\tau}\Gamma^{\tau}_{\nu\alpha}\right)\,,
\label{action}
\end{eqnarray}
where $\epsilon^{\lambda\mu\nu}=+1/\sqrt{-g}$ is the Levi-Civita tensor. The dimensionless coupling $\nu$ in front of the Chern Simons term in the action is related to the graviton mass as $\nu=\tilde\mu l/3$. For the space times which are asymptotically $\rm{AdS}_3$, the critical chiral gravity theory is at $\tilde\mu l=1$ or equivalently $\nu=1/3$ \cite{strom1}. However the warped $\rm{AdS}_3$ vacua which exhibit critical behavior at $\nu=1$ or $\tilde\mu l=3$ are also of very much interest and is recently discussed in \cite{strom2}.

The simplest non-Einsteinian solution to topologically massive gravity is called the warped $\rm{AdS}_3$ \cite{strom2}, since it involves a warped fibration. The warped $\rm{AdS}_3$ geometry is viewed as a fibration of the real line with a warp factor over $\rm{AdS}_2$. For each value of $\nu$ there exists two different solutions to the equation of motion corresponding to (\ref{action}). For spacelike, timelike \cite{nutku} and null \cite{detournay} cases there exists six such solutions. Depending on $\nu^2<1$ and $\nu^2>1$ they are called squashed and stretched solutions respectively. We are interested in studying the normal modes for a scalar field in the background of a spacelike stretched black hole which are asymptotic to warped $\rm{AdS}_3$ space time. These black holes are studied in \cite{clement1, clement2, strom2}. One of the important aspects of this geometry is that it is free from any closed timelike curves (CTC). The black hole space time is defined by the metric
\begin{eqnarray}
\nonumber ds^2=&&- N^2(r)dt^2 + l^2R^2(r)[d\phi + N_\phi(r)dt]^2 \\
 &&+\frac{l^4dr^2}{4R^2(r)N^2(r)}\,, \label{metric}
\end{eqnarray}
where
\begin{eqnarray}
R^2(r) &=& r/4\left[3(\nu^2-1)r +(\nu^2+3)(r_+ +
r_-)\right. \nonumber \\ &&\left. -4\nu\sqrt{r_+r_-(\nu^2+3)}\right]\,,\\
N^2(r) &=& \frac{(\nu^2+3)(r-r_+)(r-r_-)}{4R^2(r)}\,,\\
N_\phi(r) &=& \frac{2\nu r-\sqrt{r_+r_-(\nu^2+3)}}{2R^2(r)}\,.
\label{coeff}
\end{eqnarray}
In the above metric $r_+$ and $r_-$ are the outer and inner horizon respectively and $\nu$ is related to the warp factor as $2\nu/\sqrt{\nu^2+3}$. We will consider only the case $\nu^2>1$, since it does not contain any closed timelike curve.

The dynamics of a scalar field with mass $\mu$ in this black hole space time is given
by the Klein-Gordon equation
\begin{eqnarray}
(\square- \mu^2)\Psi(t, r, \phi) =0\,. \label{Klein}
\end{eqnarray}
Using the separation of variables
\begin{eqnarray}
\Psi(t,r,\phi)= e^{-iEt}e^{im\phi}\mathcal{U}^E (r)\,,
\end{eqnarray}
where $m \in Z$ is the azimuthal quantum number, the radial eigenvalue equation then becomes
\begin{eqnarray}
H_E\mathcal{U}^E (r)=0\,.
\end{eqnarray}
The operator $H_E$ is given by
\begin{eqnarray}
\nonumber H_E= &&\frac{d^2}{dr^2} + \frac{(2r-r_+ -r_-)}{(r-r_+)(r-r_-)}\frac{d}{dr}\\
 &&-\frac{(\alpha r^2 +\beta r +\gamma )}{(r-r_+)^2(r-r_-)^2 }\,,
\end{eqnarray}
where
\begin{eqnarray}
\alpha &=&- \frac{3E^2(\nu^2-1)-\mu^2l^2(\nu^2+3)}{(\nu^2+3)^2}\,,\\
\nonumber \beta &=& -  \frac{[E^2(\nu^2+3)(r_+ + r_-)}{(\nu^2+3)^2}\\
\nonumber &&+\frac{4\nu(E^2\sqrt{r_+r_-(\nu^2+3)}-2mE)]}{(\nu^2+3)^2}\\
&& + \frac{\mu^2l^2(r_+ +r_-)(\nu^2+3)}{(\nu^2+3)^2}\,,\\
\nonumber \gamma &=&- \frac{4m[m- E\sqrt{r_+r_-(\nu^2+3) }]}{(\nu^2+3)^2}\\
&&+ \frac{\mu^2l^2r_+r_-(\nu^2+3)}{(\nu^2+3)^2}\,.
\end{eqnarray}
In  $z= \frac{r-r_+}{r-r_-}$ coordinate, the radial operator $H_E$
can be written in the following form
\begin{eqnarray}
\nonumber H_E= z(1-z)\frac{d^2}{dz^2} + (1-z)\frac{d}{dz}
+\left(\frac{A}{z}+  \frac{B}{1-z} + C\right)\,,\\*
\label{rad1}
\end{eqnarray}
where
\begin{eqnarray}
A &=& \frac{(-2E\Omega_H^{-1}+ 2m)^2}{(r_+-r_-)^2 (\nu^2+3)^2}\,,\\
B &=&- \alpha\,,\\
C &=&- \frac{(-2E\tilde\Omega_H^{-1}+ 2m)^2}{(r_+-r_-)^2 (\nu^2+3)^2}\,,
\end{eqnarray}
and the angular velocity of the outer and inner horizons are defined as
\begin{eqnarray}
\Omega_H &=& {\frac{d\phi}{dt}}\Big\vert_{r= r_+}= -\frac{2}{2\nu r_+-\sqrt{r_+r_-(\nu^2+3)}}\,,\\
\tilde\Omega_H &=& {\frac{d\phi}{dt}}\Big\vert_{r= r_-}= -\frac{2}{2\nu r_--\sqrt{r_+r_-(\nu^2+3)}}\,.
\end{eqnarray}
Note that in the $z$ coordinate the outer horizon is situated at $z=0$ and the spatial
infinity is at $z=1$.

Next we use the ansatz
\begin{eqnarray}
\mathcal{U}^E (z) = z^{p} \left(1-z \right)^{q}
\psi(z)\,,
\end{eqnarray}
where
\begin{eqnarray}
p &=& i\sqrt{A} \nonumber \,, \\
q &=& \frac{1}{2}(1 - \sqrt{1 + 4\alpha})\,.
\end{eqnarray}
The equation corresponding to $\psi(z)$ is then given by
\begin{eqnarray}
\label{hyper} H \psi(z) = 0\,,
\end{eqnarray}
where
\begin{eqnarray}
\label{op} H = z(z-1)\frac{d^2 }{dz^2} + \{ c - (a +b +1)z \}
\frac{d}{dz} - a b
\end{eqnarray}
and
\begin{eqnarray}
a &=&  p+q +\sqrt{C}, \nonumber \\
b &=&  p+q - \sqrt{C}, \nonumber \\
c&=& 2p+1,
\end{eqnarray}

From (\ref{hyper}) and (\ref{op}) we see that the function $\psi(z)$ satisfies the
hypergeometric equation \cite{as}. The two independent solutions of $\psi$ at the outer horizon,
$r=r_+(z=0)$, are given  by
\begin{eqnarray}
\psi_{1(0)} &=& _2F_1(a,b,c,z)\,,\\
\psi_{2(0)} &=& z^{1-c}{_2F_1}(a -c+1,b-c+1,2-c,z)\,,
\end{eqnarray}
and the two independent solutions at spatial infinity, $r=\infty(z=1)$, are given by
\begin{eqnarray}
\psi_{1(1)} &=& _2F_1(a,b,a+b+1-c,1-z)\,,\\
\nonumber \psi_{2(1)} &=& (1-z)^{c-a-b}\times\\
 &&{_2F_1}(c-b,c-a,c-a-b+1,1-z)\,.
\label{infinity}
\end{eqnarray}
In order to get normal modes of the scalar field we first consider the radial solutions around spatial infinity, $z=1$, which are given by
\begin{eqnarray}
\mathcal{U}^E_{1(1)} &=& z^p(1-z)^q\psi_{1(1)}\,,\label{inf11}\\
\mathcal{U}^E_{2(1)}&=& z^p(1-z)^q\psi_{2(1)}\,.
\label{inf12}
\end{eqnarray}
Since the normal mode analysis requires square-integrability criteria for the wave function, both of the above solutions may not be useful in our case. To check the square-integrability at spatial infinity we need to get the asymptotic behavior of the solutions (\ref{inf11}) and (\ref{inf12})  which are
\begin{eqnarray}
\lim_{z\to 1}\mathcal{U}^E_{1(1)} &\simeq& (1-z)^q\,,\label{infinity11}\\
\lim_{z\to 1}\mathcal{U}^E_{2(1)} &\simeq& (1-z)^{1-q}\,.
\label{infinity1}
\end{eqnarray}
The measure  involved in the inner product of the scalar field is given by
\begin{eqnarray}
\int\sqrt{-g}dr\equiv \frac{l^3}{2}(r_+-r_-)\int\frac{dz}{(1-z)^2}.
\label{measure}
\end{eqnarray}
The square-integrability depends upon the eigenvalue $E$ of the scalar field. For the sake of simplicity of our discussion we divide the eigenvalue in two regions which are as follows
\begin{eqnarray}
|E| & <& \frac{\nu^2 +3}{2\sqrt{3}\sqrt{\nu^2-1}}\sqrt{1+\frac{4\mu^2l^2}{\nu^2+3}}\,,\label{rg1}\\
|E| & \geq& \frac{\nu^2 +3}{2\sqrt{3}\sqrt{\nu^2-1}}\sqrt{1+\frac{4\mu^2l^2}{\nu^2+3}}\,.\label{rg2}
\end{eqnarray}
Note that for the range (\ref{rg1}) only the second solution  (\ref{infinity1}) is square-integrable. But for the range (\ref{rg2}) both the solutions (\ref{infinity11}) and (\ref{infinity1}) are not square-integrable. So, from now on we will restrict ourselves in the interval (\ref{rg1}) and consider the solution  (\ref{inf12}) to analyze the normal mode in the warped $\rm{AdS}_3$ black hole background.
The analytic continuation of this solution  near
the outer horizon has the form
\begin{eqnarray}
\nonumber \tilde{\mathcal{U}}^E & =& z^{p} \left(1-z\right)^{1-q}
{_2F_1}\left(c - b, c - a , c,z\right) \\
\nonumber &+& \Theta z^{-p} \left(1-z\right)^{1-q}\times \\
&& {_2F_1}(1-a, 1-b, -c,z)\,,
\label{hor1}
\end{eqnarray}
where one can check that $|\Theta|=1$, so it should be a pure phase of the form
\begin{eqnarray}
\Theta= e^{-2i\pi\theta} = \frac{\Gamma(c- 1)\Gamma(1 - a)\Gamma(1- b)}
{\Gamma(c - b)\Gamma(c-a)\Gamma(1-c)}\,.
\end{eqnarray}
The behavior of the function $\tilde{\mathcal{U}}_E (z)$  near the outer horizon is given by
\begin{eqnarray}
\lim_{z\to 0}\tilde{\mathcal{U}}^E  \simeq z^{p} +  e^{-2i\pi\theta} z^{-p}\,.
\label{phsol1}
\end{eqnarray}
Since the radial function $\tilde{\mathcal{U}}^E$ is  singular at the outer horizon we introduce a cut off $\epsilon_H\equiv r-r_+$ and demand that the radial wave function will be zero at   $\epsilon_H$. The wave function at that point is
\begin{eqnarray}
\lim_{\epsilon_H\to 0}\tilde{\mathcal{U}}^E  = e^{p\ln\frac{\epsilon_H}{r_+-r_-} +i\pi\theta}
+  e^{-p\ln\frac{\epsilon_H}{r_+-r_-} -i\pi\theta}
\end{eqnarray}
Since $p$ is purely imaginary, by requiring that the  above equation  will be zero we get
\begin{eqnarray}
E= \Omega_Hm + \mathcal{C}(n+ 1/2-\theta)\,,
\label{eigen1}
\end{eqnarray}
where
\begin{eqnarray}
\mathcal{C}=\frac{\pi\Omega_H(r_+-r_-)(\nu^2+3)}{2\ln{\frac{\epsilon_H}{r_+-r_-}}}
\end{eqnarray}
Note that not all $m$ and  $n$  values  are allowed.
Only those values are allowed for which the constraint (\ref{rg1})  would be satisfied.
One can see that  (\ref{eigen1}) is a highly nonlinear equation for the eigenvalue, which cannot be solved analytically but must be solved numerically. This is similar to what happens in the work of  Ref. \cite{satoh} also.
Here one can check that in the limit $\nu\to 1$ the eigenvalues takes the form
\begin{eqnarray}
E= \widehat{\Omega_H}m + \widehat{\mathcal{C}}(n+ 1/2-\theta)
\end{eqnarray}
where
\begin{eqnarray}
\widehat{\Omega_H} &=&  -\frac{1}{r_+-\sqrt{r_+r_-}}\,,\\
\widehat{\mathcal{C}} &=&\frac{4\pi\widehat{\Omega_H}(r_+-r_-)}{2\ln{\frac{\epsilon_H}{r_+-r_-}}}
\end{eqnarray}

In the limit $\nu\to 1$, we can also calculate the entropy  of the scalar field for non zero $\Omega_H$ using the boundary condition at the brick wall following \cite{satoh}. The expression for the entropy is then
\begin{eqnarray}
\nonumber S_H^{\Omega_H\neq 0}(\beta) & =& \beta^2\frac{\partial F_H^{\Omega_H\neq 0}(\beta)}{\partial\beta}\\
\nonumber &\sim& \frac{d(r_+- \sqrt{r_+r_-})\ln(\frac{r_+-r_-}{\epsilon_H})}{2\pi(r_+-r_-)}\times\\
\nonumber &&\left( S^{\Omega_H\neq 0}(\beta) -N_1(N_2+1)\right)\\
&&+ N_1\sum_{m=0}^{N_2}\delta_{\theta(0,m),1/2},
\end{eqnarray}
where $N_1$ is the cutoff for the occupation number per mode introduced to regularize the grand canonical partition function, $N_2$ is the cutoff for the azimuthal quantum number, $d$ is the density of states and $F_H^{\Omega_H\neq 0}(\beta)$ is the free energy for non zero $\Omega_H$. Note, that we have written the expression in terms of the entropy $S^{\Omega_H\neq 0}$ which is given by
\begin{eqnarray}
\nonumber S^{\Omega_H\neq 0}(\beta) &&=\frac{1}{d}\left[\frac{\pi^2}{3\beta}(2N_2+1)\right.\\
\nonumber &&\left.-\Omega_H\sum_{m=1}^{N_2}m\ln(1-e^{-N_1\beta\Omega_Hm})\right.\\
\nonumber &&\left. +\frac{2}{N_1\beta}\sum_{m=1}^{N_2}\int_{x=0}^{N_1\beta\Omega_Hm}dx\ln(1-e^{-x})
\right]\\
&& + N_1(N_2+1)
\end{eqnarray}
and is calculated using the regular boundary condition at the origin $r=0$. This can be done since $r=0$ is not among the three regular singular points of the hypergeometric equation.

\section{Normal modes with generic boundary conditions}

In the previous section the solution for the normal modes of the scalar field has been obtained using  the boundary condition that the field vanishes at the brick wall. However, there could be more general boundary conditions which also lead to a well posed dynamics for the scalar field. In this section, we look for all possible boundary conditions of the scalar field so that the time evolution of the field remains unitary. This can be achieved with the help of von Neumann's method of self-adjoint extensions \cite{reed}, which is briefly described below. Consider an unbounded symmetric operator $T$ defined on dense domain $D(T)$ of a  Hilbert space ${\cal H}$.  Following von Neumann \cite{reed}, we need to find out the deficiency indices
\begin{eqnarray}
n_{\pm}(T) \equiv {\rm dim} [K_{\pm}]\,,
\end{eqnarray}
where $K_{\pm}$ are the kernel of a suitably defined operator of the form
\begin{eqnarray}
K_{\pm} \equiv {\rm Ker}(i \mp T^*)\,.
\end{eqnarray}
In terms of the deficiency indices $n_{\pm}$, the operator $T$ falls in one of the following classes :
\begin{enumerate}
\item $T$ is (essentially)
self-adjoint iff $( n_+ , n_- ) = (0,0)$.
\item $T$ has self-adjoint
extensions iff $n_+ = n_-$. In this case $K_+$ and $K_-$ are related by an unitary map which is related to the self-adjoint extension of  $T$.
\item If $n_+ \neq n_-$, then $T$ has no self-adjoint
extensions.
\end{enumerate}
In order to proceed, note that the differential form of the adjoint operator $H_E^*$
is same as that of $H_E$. However the domains of these operators need not be same. Note that the
differential operator  $H_E$ is  defined in the interval  $z\in [0, 1]$ and  is symmetric  on
the domain
\begin{eqnarray}
\nonumber D(H_E)&& \equiv \{\phi (0)  = \phi^{\prime} (0) = 0,~ \phi,~
 \phi^{\prime}~\\
\nonumber &&{\rm absolutely~ continuous},
~ \phi \in {\mathcal L}^2(\sqrt{-g}dr) \}\\*
&&
\end{eqnarray}
In order to see if the operator $H_E$ is self-adjoint in the domain $D(H_E)$, we proceed to find out its deficiency indices using the above mentioned method. We replace $E$ in the operator  $H_E$ by $+i$ and denote the obtained operator as  $H_+$.  The deficiency space solution corresponding to $E=+i$ can be obtained from
\begin{eqnarray}
H_+ \phi_+ = 0\,.
\end{eqnarray}
In general it has two independent solutions near a regular singular point. We here choose one solution
\begin{eqnarray}
\nonumber \label{phi+} \phi_+ &&= z^{p_+} \left(1-z\right)^{1-q_+}\times\\
\nonumber &&{_2F_1}\left(c_+ - b_+, c_+ - a_+ , c_+ - a_+ - b_+ + 1,
1-z\right),\\*
&&
\end{eqnarray}
which is square-integrable at spatial infinity $r=\infty$ or equivalently at $z=1$. To get the solution at the outer horizon we analytically continue  $\phi_+$ using the transformation
\begin{eqnarray}
\nonumber {_2F_1(a,b,c,z)} &=&\frac{\Gamma(c)\Gamma(c-a-b)}{\Gamma(c-a)\Gamma(c-b)}\times\\
\nonumber &&{_2F_1}(a,b,a+b-c+1,1-z)\\
\nonumber  &+& \left(1-z\right)^{c-a-b}\frac{\Gamma(c)\Gamma(a+b-c)}{\Gamma(a)\Gamma(b)}\times\\
\nonumber &&{_2F_1}(c-a,c-b,c-a-b+1,1-z)\\*
&&
\end{eqnarray}
Then the solutions at horizon up to some constant becomes
\begin{eqnarray}
\nonumber \label{phi++} \phi_+ = && e^{-i\xi_+}z^{p_+} \left(1-z\right)^{1-q_+}\times\\
\nonumber &&{_2F_1}\left(c_+ - b_+, c_+ - a_+ , c_+,z\right) \\
\nonumber &+& A_+ e^{i\xi_+}z^{-p_+} \left(1-z\right)^{1-q_+}\times \\
&&{_2F_1}(1-a_+, 1-b_+, -c_+,z)\,,
\label{hor}
\end{eqnarray}
where
\begin{eqnarray}
A_+ e^{2 i \xi_+} =
\frac{\Gamma(c_+- 1)\Gamma(1 - a_+)\Gamma(1- b_+)}
{\Gamma(c_+-b_+)\Gamma(c_+-a_+)\Gamma(1-c_+)}.
\end{eqnarray}
From
(\ref{hor}) we see that near the outer horizon, $z=0$,
$\phi_+$ behaves as
\begin{eqnarray}
\label{nh} \phi_+ \sim e^{-i\xi_+}
z^{p_+} + A_+ e^{i \xi_+} z^{-p_+},
\label{phplus}
\end{eqnarray}
The probability density obtained from (\ref{phplus}) behaves as
\begin{eqnarray}
\nonumber |\phi_+|^2 = z^{2 {\mathrm {Re}} (p_+)} + A_+e^{-2i\xi_+}z^{i2 {\mathrm {Im}}(p_+)} \\+
A_+e^{2i\xi_+}z^{-i2 {\mathrm {Im}}(p_+)}
+ A_+^2z^{-2{\mathrm {Re}} (p_+)}\,.
\label{mod}
\end{eqnarray}
where
\begin{eqnarray}
\nonumber {\mathrm{Re}} (p_+) & =& \frac{2}{\Omega_H(r_+ - r_-)(\nu^2+3)}\,, \\
{\mathrm{Im}} (p_+) & =& \frac{2m}{(r_+ - r_-)(\nu^2+3)}\,,
\end{eqnarray}
denote the real and imaginary part of $p_+$ in (\ref{mod}). Note that in
the near horizon region the measure is
\begin{eqnarray}
\sqrt{-g} dr \sim dz\,,
\end{eqnarray}
which together with (\ref{mod}) imply that  the solution $\phi_+$ is square integrable at outer
horizon if
\begin{eqnarray}
 0 < 2|{\mathrm {Re}}(p_+)| <1\,,
\label{con1}
\end{eqnarray}
Since for $E=+i$ we get only one square-integrable solution,  the deficiency index therefore is $n_+ = 1$.
Similarly for $E=-i$,  one can  show  that   $n_- = 1$. We therefore get a one parameter family of
self-adjoint extensions, which can be  parametrized by a unitary group  $e^{i\omega}, ~~ \omega \in R$ (mod $2 \pi$). To get the deficiency indices $n_\pm$ we have replaced $E=\pm i$. In general one should instead replace $E=\lambda= \lambda_R+ i\lambda_I\in \mathbb{C}$ and its complex conjugate  to find out the deficiency indices $n_\pm$. In that case the expressions for $\rm{Re}(p_+)$ will be given by
\begin{eqnarray}
\rm{Re}(p_+)&=&\frac{2\Omega_H^{-1}\lambda_I}{(r_+-r_-)(\nu^2+3)}\,.
\end{eqnarray}
From the above equations we can see that the square-integrability of $\phi_{\pm}$ is ensured only in a band like region in the complex $\lambda$-plane.
\begin{figure}
\includegraphics[width=0.40\textwidth, height=0.15\textheight]{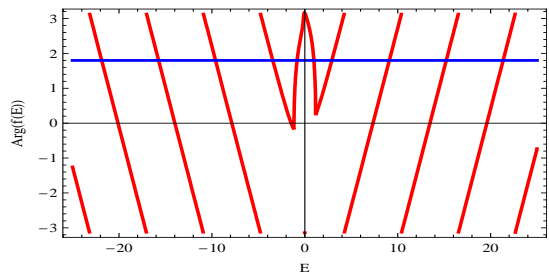}
\caption {A schematic plot of argument of $f(E)$ with $E$ for massless scalar field. The red lines correspond to the argument of $f(E)$, while the blue line corresponds to a constant value of right hand side of (\ref{f(E)}). The intersections between red and blue lines refer to the energy eigenvalues of the scalar field.}
\label{fig1}
\end{figure}
Using von Neuman's analysis, it is possible to construct the domain
in which the operator $H_E$ would be self-adjoint when in the interval (\ref{con1}).
It is given by
\begin{eqnarray}
D_\omega(H_E) = D(H_E) \oplus \Sigma \{ \phi_+ + e^{i\omega} \phi_- \},
\end{eqnarray}
where $\Sigma$ is an arbitrary complex constant. As the domain or the boundary conditions contain the self-adjoint extension parameter $\omega$, it is expected that the spectrum and the partition function would also be a function of $\omega$. However, in the present case, it is not
clear as to how to fit the solution of the physical problem with real
eigenvalues to this domain. We shall therefore pursue another approach
to obtain the physical solutions of the problem. Note that near the
outer horizon the two linearly independent solutions are given by
\begin{eqnarray}
\mathcal{U}^E_{1(0)} &=& z^p(1-z)^q\psi_{1(0)}\,,\label{inf011}\\
\mathcal{U}^E_{2(0)} &=& z^p(1-z)^q\psi_{2(0)}\,.
\label{inf012}
\end{eqnarray}
Therefore the most general solution would be the linear sum of the two
\begin{eqnarray}
\mathcal{U}=\mathcal{U}^E_{1(1)} +
D\mathcal{U}^E_{2(1)} \,.
\label{gsol}
\end{eqnarray}
The behavior of the function $\mathcal{U}^E (z)$  near the horizon is given by
\begin{eqnarray}
\lim_{z\to 0}\mathcal{U}^E (z)  \simeq z^p + D z^{-p}\,.
\label{phsol2}
\end{eqnarray}
Expression (\ref{phsol1}) and (\ref{phsol2})
physically should denote the same quantity, namely the behavior of the
scalar field near the outer horizon, and they would be compatible if the coefficient of
$z^{-p}$ are identical.  The eigenvalue can be determined from the equation
\begin{eqnarray}
f(E)\equiv e^{-2i\pi\theta}= D\,.
\label{f(E)}
\end{eqnarray}
The above equation can not in general  be solved analytically. One can plot the two different sides of the equation and from the intersections of the plots the eigenvalues can be found out as shown in FIG. \ref{fig1}.

\section{Discussion}

In this paper we have considered the scalar field propagation in the background of a warped $\rm{AdS}_3$ black hole, which arises in topologically massive gravity. We have started our analysis with the conventional boundary conditions where the scalar field vanishes at a brick wall cutoff near the outer horizon and at spatial infinity. Unlike the BTZ case, the scalar field admits normalizable states only for a certain range of the energy eigenvalues of the Klein-Gordon operator. In the limit when $\nu \rightarrow 1$, this restriction on the energy eigenvalue disappears. This is consistent with the corresponding result for the BTZ, although a direct comparison between the two is difficult. In the limit $\nu\to 1$, we have also calculated the entropy of the system using the method of \cite{satoh}.

As the physical results depend directly on the choice of boundary conditions, it is useful to investigate what other boundary conditions are possible consistent with the unitary time evolution of the system. This analysis has been performed using the self-adjoint extension technique proposed by von Neumann. We have shown that for certain ranges of the system parameter, the radial part of the Klein-Gordon operator admits a one parameter family of self-adjoint extensions. There is however some subtlety associated with this procedure due to the existence of a band like region where the square integrability of the scalar field can be demanded and self adjoint extension exists. Here we propose a physically motivated different procedure \cite{wilc}. This is similar to the corresponding analysis for the case of BTZ black hole \cite{cgg}, although the full implication of such a result needs further investigation.

The algebraic analysis of the near-horizon conformal structure by some of the present authors \cite{ksg1,ksg2,ksg3} using the self-adjoint extension of the Klein-Gordon operator led to the existence of the Virasoro algebra in the near-horizon region, which was consistent with the black hole entropy. It is plausible that a similar analysis using the self-adjoint extension discussed in this paper would be consistent with the conjecture regarding the existence of two copies of the Virasoro algebra  as a symmetry of the warped $\rm{AdS}_3$ black hole \cite{strom2}.

\end{document}